\documentclass[12pt]{article}

\begin{document}
\begin{titlepage}
\begin{center}

\vskip 1cm


{\large \bf {Minimal Flavour Violation and Multi-Higgs Models}}

\vskip 1cm

F. J. Botella  $^a$ \footnote{fbotella@uv.es}, 
G. C. Branco  $^b$ \footnote{gustavo.branco@cern.ch and 
gbranco@ist.utl.pt}, and M. N.
Rebelo $^b$ \footnote{margarida.rebelo@cern.ch and
rebelo@ist.utl.pt}

\vspace{1.0cm}

{\it $^a$ Departament de F\' \i sica Te\`orica and IFIC,
Universitat de Val\`encia-CSIC, E-46100, Burjassot, Spain.} \\
{\it $^b$ Departamento de F\'\i sica and Centro de F\' \i sica Te\' orica
de Part\' \i culas (CFTP),
Instituto Superior T\' ecnico, Av. Rovisco Pais, P-1049-001 Lisboa,
Portugal.}

\end{center}

\vskip 3cm

\begin{abstract}
We propose an extension of the hypothesis of Minimal Flavour 
Violation (MFV) to general multi-Higgs Models without 
the assumption of Natural Flavour Conservation in the Higgs 
sector. We study in detail under what conditions the neutral 
Higgs couplings are only functions of $V_{CKM}$ and propose 
a MFV expansion for the neutral Higgs couplings to fermions.
\end{abstract}

\end{titlepage}

\newpage

\section{Introduction}
The Standard Model (SM) of the electroweak and strong interactions
has had an impressive success in accounting for most of the presently
available experimental data. The discovery of non-vanishing 
neutrino masses provided a notable exception \cite{Amsler:2008zz}, 
pointing towards
New Physics (NP), since in the SM neutrinos are strictly massless.

In spite of its great success, there is a general consensus that 
the SM including its simple extension incorporating neutrino masses,
cannot be the ``final theory''. One of the reasons for this, has to 
do with the large number of free parameters, most of them arising 
from the flavour sector of the SM. This proliferation
of free parameters reflects the fact that the flavour
structure of Yukawa couplings is not constrained by gauge invariance.  
In the SM, flavour changing neutral currents (FCNC) are forbidden
at tree level both in the gauge and the Higgs sectors. From the
early stages of gauge theories, some principles of flavour
conservation by neutral currents have been introduced both in 
the gauge sector through a generalization of the GIM mechanism 
\cite{Glashow:1970gm},
as well as in the scalar sector through the principle of Natural 
Flavour Conservation (NFC) proposed by Glashow and Weinberg 
\cite{Glashow:1976nt}.
It is interesting to note that one may have non-zero but naturally 
suppressed FCNC in the gauge sector in models where vector-like 
quarks \cite{vector1} \cite{Bento:1991ez}, \cite{vector2}
are added to the SM. In this case, gauge mediated FCNC 
arise at tree level, suppressed by the small ratio $m^2/M^2$
where $m$ and $M$ denote standard quark masses and 
vector-like quark masses, respectively. Vector-like quarks arise 
in various extensions of the SM, including $E_6$
grand-unified theories and extra-dimension models. Other 
motivations for considering vector-like quarks  include the 
possibility of finding a solution to the strong CP 
problem \cite{Barrnelson}, \cite{Bento:1991ez}
and accounting for \cite{vector2} the potentially large CP asymmetry
recently observed in $B_s \rightarrow J/\Psi \phi $ decays 
\cite{:2008fj}, \cite{Aaltonen:2007he}. Recently, a different possibility
was considered \cite{Pich:2009sp} to avoid tree-level FCNC processes in the 
framework of two Higgs doublet models, allowing for new sources of CP violation.\\

All the flavour changing transitions in 
the SM are mediated by charged weak currents
with the flavour mixing controlled by the Cabibbo-Kobayashi-Maskawa (CKM)
matrix, $V_{CKM}$
\cite{ckm}.  
Any extension of the SM which attempts
at solving the flavour puzzle has to confront the
strict limits on FCNC processes
as well as limits on CP violating transitions leading, for example, to
electric dipole moments of quarks and leptons \cite{Raidal:2008jk}.

In the scalar sector, it has been considered the possibility of 
allowing for deviations of strict NFC by invoking the
presence of suppression factors \cite{Ant+hall} \cite{Joshipura:1990pi} 
involving small off-diagonal 
elements of the quark mixing matrix $V_{CKM}$. The first 
models of this type were proposed by Branco, Grimus and Lavoura 
(BGL) \cite{Branco:1996bq} who have shown that there are extensions of the SM 
with two Higgs doublets and an additional discrete symmetry, 
where there are FCNC at tree level, with couplings entirely 
determined in terms of the CKM matrix elements, with no other 
free parameters. In some variants of these models \cite{Branco:1996bq}
the Higgs particles can be relatively light, 
without entering in conflict with the stringent limits on FCNC 
processes.

The success of the SM and its CKM mechanism of mixing and CP 
violation shows that if there are New Physics contributions
to flavour changing interactions at the Tev scale its couplings 
should occur at a much higher scale or else should be 
strongly non-generic. This is natural and in
a certain sense to be expected if one takes into account that 
flavour changing transitions in the SM have a special flavour 
structure, not predicted within its framework. 
For example, in the SM there is no explanation for the pattern
of flavour mixings and in particular why 
$(V_{CKM})_{12} \sim (m_d/m_s)^{1/2} $ while 
$(V_{CKM})_{23} \sim (m_s/m_b) $. 

One of the suggestions for the flavour structure of New Physics 
is the proposal of Minimal Flavour Violation (MFV) \cite{Buras:2000dm}, 
\cite{D'Ambrosio:2002ex} where all new flavour changing transitions are 
controlled by the CKM matrix. The gauge sector of the Standard
Model (SM) with three generations
of quarks and leptons has a large $G_F = U(3)^5$ flavour symmetry
which is only broken by Yukawa couplings. One may formally recover 
 \cite{D'Ambrosio:2002ex} this flavour symmetry by 
promoting Yukawa couplings to auxiliary fields $Y$, transforming
under  $G_F$ in such a way that Yukawa interactions become
$G_F$ invariant. Then an effective theory arising from New Physics
is of MFV type  if all higher order operators,
constructed from SM fields and $Y$ fields are formally invariant 
under $G_F$. This hypothesis, together with the realization that in the SM 
Yukawa couplings for all fermions, except the top, are small,
leads to specific predictions 
\cite{phenmvf}.

If one regards the SM as an effective theory, valid up to
some energy scale $\Lambda$, then in order to have a solution of the
hierarchy problem, one expects the scale $\Lambda$
of New Physics to be of the order of a few TeV. The above considerations
have motivated the idea of Minimal Flavour Violation (MFV) both in the
quark \cite{Buras:2000dm}, \cite{D'Ambrosio:2002ex} and lepton sectors 
\cite{Cirigliano:2005ck}, \cite{Branco:2006hz}. 
The MFV hypothesis requires that all 
flavour and CP violating interactions be related to the structure of 
Yukawa couplings and controlled by $V_{CKM}$.

The MFV idea has been applied to two Higgs doublet extensions of the SM
where there is Natural Flavour Conservation (NFC) in the Higgs sector
at tree level, as it is the case in the minimal supersymmetric extension
of the Standard Model (MSSM). \\

In this paper we examine how to extend the ideia of Minimal Flavour 
Violation to the scalar sector with two and three Higgs doublets, 
without the assumption of Natural Flavour Conservation in the 
Higgs sector. This paper is organized as follows:
in section 2 we recall  the
important requirement of rephasing invariance, and in section 3 we analyse
in detail how the requirement of MFV can be fulfilled in the
context of an extension of the SM where two Higgs doublets
are  introduced. In section 4 we propose a general MFV expansion
of the neutral Higgs couplings to quarks and we stress the
important role of discrete symmetries in fixing the parameters of
this expansion. The case of three Higgs
doublets in the context of MFV is analysed in section 5 and 
our conclusions are presented
in the last section.

\section{The Requirement of Rephasing Invariance}

As we have seen, the definition of MFV includes the requirement
that all flavour transitions are controlled by the CKM matrix.
Let us consider a FCNC transition connecting, for definiteness, a
$Q= - 1/3$ quark $d_j$ to a different quark of the same charge 
$d_k$. The transition could be mediated by a scalar or a vector
boson:
\begin{eqnarray}
{\cal L}_{scalar}  &=& \overline{d_{Lj}} \ {\Gamma^S}_{jk} \ \ d_{Rk} \ S \\
\label{lhc}
{\cal L}_{vector}  &=& \overline{d_{Lj}} \ 
{\Gamma^V}_{jk} \ \gamma_\mu \ d_{Lk} \ V^{\mu}
\label{lep}
\end{eqnarray}
Note that the couplings $\Gamma^S$, $\Gamma^V$ may arise at 
tree level or in higher orders. Let us assume that the quark 
mass matrices have been diagonalized, so that $d_j$ denote quark mass 
eigenstates. Under rephasing of the quark fields:
\begin{equation}
d_j \rightarrow d^{\prime}_j = \exp{(-i \beta_j)}\  d_j
\end{equation}
the couplings ${\Gamma^S}_{jk}$ and ${\Gamma^V}_{jk}$ have to 
transform in such a way that the interactions of Eqs.~(\ref{lhc})
and (\ref{lep})
remain rephasing invariant. This implies that under rephasing
\begin{equation}
\Gamma_{jk} \rightarrow \Gamma^{\prime}_{jk} =  
\exp{[i (\beta_k - \beta_j)]}\ \Gamma_{jk} 
\end{equation}
The fact that in MFV theories, the flavour dependence of
$\Gamma_{jk}$ is completely controlled by the CKM matrix, 
severely restricts the functional dependence of $\Gamma_{jk}$
on $V_{CKM}$. The simplest forms allowed by rephasing 
invariance are:
\begin{equation}
\Gamma_{jk} = \sum_\alpha c_\alpha V_{\alpha j}  V^*_{\alpha k}
\label{abc}
\end{equation}
where $c_{\alpha}$ are rephasing invariant coefficients. 
In the sequel, we shall see that the simplest 
two Higgs doublet (2HD) models which conform to the MFV requirement
do have FCNC couplings with such functional dependence on   $V_{CKM}$.

\section{The case of Two Higgs Doublets}
In this section, we analyse in detail how the requirement of 
MFV can be fulfilled in the context of an extension of the SM,
where two Higgs doublets are introduced. In order to fix our notation,
we explicitly write the Yukawa interactions:
\begin{equation}
L_Y = - \overline{Q^0_L} \ \Gamma_1 \Phi_1 d^0_R - \overline{Q^0_L}\  
\Gamma_2 \Phi_2 d^0_R - \overline{Q^0_L} \ \Delta_1 \tilde{ \Phi_1} u^0_R - 
\overline{Q^0_L} \ \Delta_2 \tilde{\Phi_2} u^0_R \ + \mbox{h. c.}
\label{1e2}
\end{equation} 
where $\Gamma_i$ and $\Delta_i$ denote the Yukawa couplings of the 
lefthanded quark doublets $Q^0_L$ to the righthanded quarks $d^0_R$,
$u^0_R$ and the Higgs doublets $\Phi_j$. The quark mass matrices generated
after spontaneous gauge symmetry breaking are given by:
\begin{eqnarray}
M_d = \frac{1}{\sqrt{2}} ( v_1  \Gamma_1 +
                           v_2 e^{i \alpha} \Gamma_2 ), \quad 
M_u = \frac{1}{\sqrt{2}} ( v_1  \Delta_1 +
                           v_2 e^{-i \alpha} \Delta_2 ),
\label{mmmm}
\end{eqnarray}
where $v_i \equiv |<0|\phi^0_i|0>|$ and $\alpha$ denotes the relative phase 
of the vacuum expectation values (vevs) of the neutral components of 
$\Phi_i$. The matrices $M_d, M_u$ are diagonalized by the usual 
bi-unitary transformations:
\begin{eqnarray}
U^\dagger_{dL} M_d U_{dR} = D_d \equiv {\mbox diag}\ (m_d, m_s, m_b) 
\label{umu}\\
U^\dagger_{uL} M_u U_{uR} = D_u \equiv {\mbox diag}\ (m_u, m_c, m_t)
\label{uct}
\end{eqnarray}
In terms of the quark mass eigenstates $u, d$, the Yukawa couplings are:
\begin{eqnarray}
L_Y & = & \frac{\sqrt{2} H^+}{v} \bar{u} \left(
V N_d \gamma_R + N^\dagger_u \ V \gamma_L \right) d +  \mbox{h.c.} - 
\frac{H^0}{v} \left(  \bar{u} D_u u + \bar{d} D_d \ d \right) - 
\nonumber \\
& - & \frac{R}{v} \left[\bar{u}(N_u \gamma_R + N^\dagger_u \gamma_L)u+
\bar{d}(N_d \gamma_R + N^\dagger_d \gamma_L)\ d \right] + \\
& + & i  \frac{I}{v}  \left[\bar{u}(N_u \gamma_R - N^\dagger_u \gamma_L)u-
\bar{d}(N_d \gamma_R - N^\dagger_d \gamma_L)\ d \right]
\nonumber
\end{eqnarray}
where $v \equiv \sqrt{v_1^2 + v_2^2} = (\sqrt{2} G_F)^{-1/2} \approx 
\mbox{246 GeV}$,  $G_F$ is the Fermi coupling constant,  
$\gamma_L = (1 - \gamma_5)/2$,
$\gamma_R = (1 + \gamma_5)/2$, $V$ stands for the $V_{CKM}$ matrix
and $H^0$, $R$ are orthogonal combinations of the fields  $\rho_j$,
arising when one expands  \cite{Lee:1973iz} the neutral scalar fields
around their vevs, $ \phi^0_j =  \frac{e^{i \alpha_j}}{\sqrt{2}} 
(v_j + \rho_j + i \eta_j)$. Similarly, $I$ denotes the linear combination
of $\eta_{j}$ orthogonal to the neutral Goldstone boson. The 
physical neutral Higgs fields are combinations of 
$H^0$, $R$ and $I$.

The Flavour Changing Neutral Yukawa Couplings (FCNYC) are controlled
by the matrices $N_d$,  $N_u$, given by:
\begin{eqnarray}
N_d = \frac{1}{\sqrt{2}} U^\dagger_{dL}\  ( v_2  \Gamma_1 -
                           v_1 e^{i \alpha} \Gamma_2 )\  U_{dR}, 
\label{ndnd} \\
N_u = \frac{1}{\sqrt{2}}U^\dagger_{uL} ( v_2  \Delta_1 -
\label{nunu}                           v_1 e^{-i \alpha} \Delta_2 )\  U_{uR}
\end{eqnarray}
For generic two Higgs doublet models, the coupling matrices 
$N_d$, $N_u$ are non-diagonal and arbitrary. We are interested in 
analysing under what circunstances the flavour structure of 
$N_d$, $N_u$  is entirely controlled by the CKM matrix, as required
by the MFV paradigm. 

For definiteness, let us consider
$N_d$, which can be written \cite{Lavoura:1994ty} 
from Eqs.(\ref{mmmm}), (\ref{umu})and (\ref{ndnd}) :
\begin{equation}
N_d = \frac{v_2}{v_1} D_d - \frac{v_2}{\sqrt{2}}
\left( \frac{v_2}{v_1} +  \frac{v_1}{v_2}\right)  U^\dagger_{dL}
e^{i \alpha} \Gamma_2 \ U_{dR}
\label{nd}
\end{equation}
From Eq.~(\ref{nd}), one sees that there are two obstacles which
one has to surmount in order to have $N_d$ entirely controlled by $V_{CKM}$: \\

(i)  It is $U_{dL}$ rather than the combination $U^\dagger_{uL}  U_{dL}$ 
corresponding to  $V_{CKM}$ that appears in $N_d$ given by  Eq.~(\ref{nd}) \\

(ii) How to get rid of the dependence on  $ U_{dR}$? \\

The first difficulty can be solved by means of a flavour 
symmetry constraining  $U_{uL}$ to have mixing only among
two generations, for example:
\begin{eqnarray}
U_{uL}  = \left[\begin{array}{ccc}  
\times  & \times & 0 \\
\times & \times & 0  \\
0 & 0 & 1 
\end{array}\right] 
\label{calm}
\end{eqnarray}
In this case one has:
\begin{equation}
(V_{CKM})_{3j} = (U_{dL})_{3j}
\label{3j}
\end{equation}
In order to surmount obstacle (i) one has to further
require that the above symmetry  
should also impose that the dependence of the second term of 
Eq.~(\ref{nd}) on
$U_{dL}$ be only on elements of its third row, $(U_{dL})_{3j}$.
We now turn to question (ii) namely, how
to avoid the dependence on $U_{dR}$. Let us assume that the flavour
structures o  $\Gamma_2$, is such that:
\begin{equation}
\Gamma_2 \propto P M_d
\label{pro}
\end{equation}
Where $P$ is a fixed matrix. In this case:
\begin{equation}
 U^\dagger_{dL} \Gamma_2 \ U_{dR} \propto  
U^\dagger_{dL} P M_d \  U_{dR}   \propto U^\dagger_{dL} P\   U_{dL} D_d
\end{equation}
thus answering question (ii).

Let us now see what should be the flavour structure of $\Gamma_1$,
$\Gamma_2$ so that a fixed matrix P exists, statisfying Eq.~(\ref{pro}).
One way of achieving this is by having
\begin{eqnarray}
P \Gamma_2 = k \Gamma_2 \label{pg2} \\
P \Gamma_1 = 0 \label{pg1} 
\end{eqnarray}
where $k$ is a constant.

Branco, Grimus and Lavoura  have shown \cite{Branco:1996bq}
that it is possible
to find a symmetry which, when imposed to a two Higgs doublet
extension of the SM, leads to a structure for  $\Gamma_i$ and 
$\Delta_i$ such that there are scalar FCNC at tree level, with
strength completely controlled by $V_{CKM}$. BGL have imposed 
the following symmetry $S$ on the Lagrangian:
\begin{equation}
Q^0_{L3} \rightarrow \exp{(i \alpha)}\  Q^0_{L3}, \qquad
u^0_{R3} \rightarrow \exp{(i 2\alpha)} u^0_{R3}, \qquad
\Phi_2   \rightarrow \exp{(i \alpha)} \Phi_2 \label{bgl}
\end{equation}
where $\alpha \neq 0, \pi$, with all other fields transforming
trivially under $S$. The most general Yukawa couplings consistent
with this symmetry have the following structure:
\begin{eqnarray}
 \Gamma_1 & = & \left[\begin{array}{ccc}  
\times  & \times & \times \\
\times & \times &  \times \\
0 & 0 & 0 
\end{array}\right]; \qquad
 \Gamma_2   =  \left[\begin{array}{ccc}  
0 & 0 & 0  \\
0 & 0 & 0 \\
\times & \times &  \times 
\end{array}\right] \label{gam}\\
 \Delta_1  & = & \left[\begin{array}{ccc}  
\times  & \times & 0 \\
\times & \times &  0 \\
0 & 0 & 0 
\end{array}\right]; \qquad 
 \Delta_2   =  \left[\begin{array}{ccc}  
0  & 0 & 0 \\
0 & 0 &  0 \\
0 & 0 & \times
\end{array}\right] \label{del}
\end{eqnarray}
where $\times$ denotes an arbitrary entry while the zeros are imposed
by the symmetry $S$.

It is clear that these Yukawa couplings guarantee that Eqs.~(\ref{calm})
and (\ref{3j}) are satisfied. They also satisfy Eqs.~(\ref{pro}), 
(\ref{pg2}), (\ref{pg1}) with 
\begin{eqnarray}
P  = \left[\begin{array}{ccc}  
0  & 0 & 0 \\
0 & 0 & 0  \\
0 & 0 & 1 
\end{array}\right];\ \ \     
\frac{v_2}{\sqrt{2}} e^{i \alpha} \Gamma_2 = P  M_d; \ \ \ k=1
\label{pgk}
\end{eqnarray}
It follows then that the Yukawa couplings of Eqs.~(\ref{gam}) and
(\ref{del}) lead to FCNC at tree level, entirely determined by $V_{CKM}$.
Notice that in this example there are no Higgs mediated FCNC
in the up sector, which is due to the fact that the $\Delta_i$
matrices are block diagonal with each one of these
matrices having non-zero entries in different blocks. This also  
automatically leads to a matrix $U_{uL}$ which is block diagonal and therefore 
of the form given by Eq.~(\ref{calm}). The structure of zeros 
in the matrix $\Gamma_2$ leads to the important relation:
\begin{equation}
\left( U^\dagger_{dL} \Gamma_2 \right)_{ij}=   
(U^\dagger_{dL})_{i3}  (\Gamma_2)_{3j} =
(V^\dagger_{CKM})_{i3}  (\Gamma_2)_{3j}
\end{equation}
this result together with Eqs.~(\ref{pro}),(\ref{pg2}), (\ref{pg1}) 
and (\ref{nd}) leads to $N_d$ given by   \cite{Branco:1996bq}
\begin{equation}
(N_d)_{ij} = \frac{v_2}{v_1} (D_d)_{ij} - 
\left( \frac{v_2}{v_1} +  \frac{v_1}{v_2}\right) 
(V^\dagger_{CKM})_{i3} (V_{CKM})_{3j} (D_d)_{jj} \label{24}
\end{equation}
whereas
\begin{equation}
N_u = - \frac{v_1}{v_2} \mbox{diag} \ (0, 0, m_t) +  \frac{v_2}{v_1}
\mbox{diag} \ (m_u, m_c, 0) \label{25}
\end{equation}
In this example, the Higgs mediated FCNC are suppressed by the third 
row of the matrix  $V_{CKM}$ and have the structure of Eq.~(\ref{abc})
A crucial feature in this example is the fact that each row of $M_d$
only receives contribution from a single Higgs field and the same 
applies to $M_u$.

The example given above corresponds to a class of six different models,
as was emphasized in \cite{Branco:1996bq} .
Three of these models have  FCNC only in the down sector, 
and are obtained from the three different
projection matrices of a form similar to P in Eq.~(\ref{pgk}), 
the other two cases with the diagonal entry
in the other two possible entries. In these 
additional cases the suppression
in the Higgs mediated FCNC is  not as large as that of the example given
above. Another three models
are obtained by exchanging the the patterns of zeros of $\Gamma_i$ matrices
with $\Delta_i$ matrices, leading to FCNC in the up sector, and
flavour conservation in the down sector.

\section{MFV Expansion of Yukawa Couplings}
The  neutral Higgs interactions, beyond those present in the SM,
i.e., couplings to $R$ and $I$, are those that may introduce Higgs
mediated FCNC and are given by Eqs.~(\ref{nunu}), where $N_d$ and $N_u$
are written in the quark mass eigenstate basis. In a weak basis these
couplings are:
\begin{eqnarray}
N^0_d =  U_{dL} \ N_d \ U^\dagger_{dR}=
\frac{1}{\sqrt{2}} \  ( v_2  \Gamma_1 -
                           v_1 e^{i \alpha} \Gamma_2 ), \\
N^0_u =  U_{uL} \ N_u \ U^\dagger_{uR}=
\frac{1}{\sqrt{2}} \  ( v_2  \Delta _1 -
                           v_1 e^{i \alpha} \Delta _2 )
\end{eqnarray}
All other couplings involving neutral scalars are flavour conserving,
therefore they are not relevant for our analysis. The question
that we address in this section is how to find a general
expansion of $N^0_d$, $N^0_u$ which conforms to the MFV requirements.
It is clear that a necessary condition for $N^0_d$, $N^0_u$ to 
be of the MFV type is that they should be functions of $M_d$, $M_u$ 
and no other flavour dependent couplings. The terms entering
in the expansion of $N^0_d$, $N^0_u$ should have the right 
transformation properties under weak basis (WB) transformations,
defined by:
\begin{equation}
Q^0_L \rightarrow W_L \  Q^0_L, \qquad
d^0_R \rightarrow W^d_R \ d^0_R, \qquad
u^0_R \rightarrow  W^u_R \ u^0_R \qquad 
\label{wea}
\end{equation}
Under a WB transformation defined by Eq.~(\ref{wea}), the 
quark mass matrices $M_d$,  $M_u$ transform as:
\begin{equation}
M_d \rightarrow W^\dagger_L \  M_d \ W^d_R; \qquad
M_u \rightarrow W^\dagger_L \  M_u \ W^u_R
\end{equation}
The matrices $U_{dL}$, $U_{dR}$, $U_{uL}$, $U_{uR}$ defined in 
Eqs.~(\ref{umu}), (\ref{uct}) transform under a WB transformation 
in the following way:
\begin{eqnarray} 
U_{dL} \rightarrow W^\dagger_L \ U_{dL}; \ \ \ 
U_{uL} \rightarrow W^\dagger_L \ U_{uL}; \ \ \ 
U_{dR} \rightarrow W^{d \dagger}_R \ U_{dR}; \ \ \ 
U_{uR} \rightarrow W^{u \dagger}_R \ U_{uR}
\end{eqnarray}
The Hermitian matrices $H_d$,  $H_u$ with $H_{d,u} \equiv  (M_{d,u}) 
(M^\dagger_{d,u})$
transform under a WB transformation as: 
\begin{equation}
H_d \rightarrow W^\dagger_L \  H_d \ W_L; \qquad
H_u \rightarrow W^\dagger_L \  H_u \ W_L
\end{equation}
From Eqs.~(\ref{umu}), (\ref{uct}) it follows that:
\begin{equation}
U^\dagger_{dL} \ H_d U_{dL} = D^2_d
\end{equation}
with analogous expression for $H_u$. It is convenient to write $H_d$,  $H_u$ in 
terms of projection operators \cite{Botella:2004ks}:
\begin{equation}
H_d = \sum_i {m^2_{d}}_i P^{dL}_i
\end{equation}
where:
\begin{equation}
P^{dL}_i =  U_{dL}P_i U^\dagger_{dL}
\end{equation}
with 
\begin{equation}
(P_i)_{jk} = \delta_{ij} \delta_{ik}
\end{equation}
Obviously, analogous expressions hold for $H_u$. It is clear that under a WB 
transformation, $N^0_d$,  $N^0_u$ transform as $M_d$,  $M_u$. A MFV expansion for
$N^0_d$,  $N^0_u$ with proper transformation properties under a WB transformation
can then be built with terms proportional to  $M_d \ (M_u)$ respectively, 
as well as products
of terms transforming as  $H_d$ and  $H_u$ multiplying  $M_d \ (M_u)$ 
respectively:
\begin{eqnarray}
N^0_d = \lambda_1 \ M_d + \lambda_{2i} \  U_{dL}P_i U^\dagger_{dL} \ M_d +
\lambda_{3i} \  U_{uL}P_i U^\dagger_{uL} \ M_d + ... \\
N^0_u =  \tau_1 \ M_u + \tau_{2i} \  U_{uL}P_i U^\dagger_{uL} \ M_u +
\tau_{3i} \  U_{dL}P_i U^\dagger_{dL} \ M_u + ...
\end{eqnarray}
In the  quark mass eigenstate basis $N^0_d$, $N^0_u$ become:
\begin{eqnarray}
N_d = \lambda_1 \ D_d + \lambda_{2i} \ P_i  \ D_d + 
\lambda_{3i} \ (V_{CKM})^\dagger \ P_i \ V_{CKM} \ D_d + ...  
\label{38} \\
N_u =  \tau_1 \ D_u + \tau_{2i} \ P_i  \ D_u  + 
\tau_{3i} \ V_{CKM} \  P_i \ (V_{CKM})^\dagger \  D_u  + ... \label{39} 
\end{eqnarray}
which conforms explicitly to the MFV requirement.
Terms of the form $U_{dL}P_i U^\dagger_{dL} \ M_d$ and 
$ U_{uL}P_i U^\dagger_{uL} \ M_u$
do not lead to Higgs mediated FCNC, whereas terms of the form 
$ U_{uL}P_i U^\dagger_{uL} \ M_d$
and  $U_{dL}P_i U^\dagger_{dL} \ M_u$ do lead to FCNC.
At this stage the lambda and tau coefficients of these expansions
appear as free parameters. This was to be expected, since the expansions 
of Eqs.~(\ref{38}), (\ref{39}), conform to the
MFV requirement but have no further restriction.
In theories where the MFV requirement
results from the imposition of a symmetry on the Lagrangian, the
coefficients lambda and tau are constrained.\\

Comparing Eqs.~(\ref{24}) and (\ref{25}) to Eqs.~(\ref{38}) 
and (\ref{39}) one realizes that the BGL example presented in the 
previous section corresponds to the following truncation of 
our MFV expansion:
\begin{eqnarray}
N^0_d = \frac{v_2}{v_1} M_d - \left( \frac{v_2}{v_1} +  
\frac{v_1}{v_2}\right)   U_{uL}P_3 U^\dagger_{uL} \ M_d 
\label{bgl1} \\
N^0_u = \frac{v_2}{v_1} M_u - 
\left( \frac{v_2}{v_1} +  \frac{v_1}{v_2}\right)  
 U_{uL}P_3 U^\dagger_{uL} \ M_u \label{bgl2}
\end{eqnarray}
This result, together with equations:
\begin{eqnarray}
N^0_d = \frac{v_2}{v_1} M_d - \frac{v_2}{\sqrt{2}}
\left( \frac{v_2}{v_1} +  \frac{v_1}{v_2}\right)  
e^{i \alpha} \Gamma_2 \\
N^0_u = \frac{v_2}{v_1} M_u - \frac{v_2}{\sqrt{2}}
\left( \frac{v_2}{v_1} +  \frac{v_1}{v_2}\right)  
e^{-i \alpha} \Delta_2
\end{eqnarray}
implies that the BGL model is fully defined in a covariant way
under WB transformations by:
\begin{eqnarray}
\frac{v_2}{\sqrt{2}} e^{i \alpha} \Gamma_2 =  
U_{uL}P_3 U^\dagger_{uL} \ M_d  \label{lala} \\
\frac{v_2}{\sqrt{2}} e^{- i \alpha} \Delta_2 =  
U_{uL}P_3 U^\dagger_{uL} \ M_u 
\end{eqnarray}
The factors multiplying $\Gamma_2$ and $\Delta_2$ coincide with the 
coefficients for these matrices in the expressions of $M_d$ and $M_u$.
Replacing in these equations the mass matrices written in terms
of the Yukawa couplings one obtains: 
\begin{eqnarray}
U_{uL}P_3 U^\dagger_{uL}\Gamma_2 = \Gamma_2; \qquad 
U_{uL}P_3 U^\dagger_{uL}\Gamma_1 = 0  \label{g1g2} \\
U_{uL}P_3 U^\dagger_{uL}\Delta_2 = \Delta_2; \qquad
U_{uL}P_3 U^\dagger_{uL}\Delta_1 = 0  \label{d1d2} 
\end{eqnarray}
These relations are the generalization to an arbitrary basis
of the relations satisfied by  the BGL model, namely
$P_3 \Gamma_2 = \Gamma_2$, $P_3 \Gamma_1 = 0$, 
$P_3 \Delta_2 = \Delta_2$ and $P_3 \Delta_1 = 0$ which result from the
imposed symmetry. Now, we show, that in fact, in this case there
is a WB where the matrices $ \Gamma_1$, $ \Gamma_2$, $\Delta_1$ and
$\Delta_2$ have the forms given by   Eqs.~(\ref{gam}) 
and (\ref{del}). Starting from a WB where $M_u$ is real and diagonal, 
and therefore $U_{uL}=1$, 
we may perform a WB 
transformation by choosing $W_L$ and $W^u_R$ block diagonal with mixing 
in the (12) block only. As a result, the matrix $M_u$ will also be 
block diagonal, in this WB.
Eq.~(\ref{lala}) becomes:
\begin{equation}
\frac{v_2}{\sqrt{2}} e^{i \alpha} \Gamma_2 =
W^\dagger_L(12) \ P_3 W_L(12)\ M_d =  P_3 \  M_d
\end{equation}
which is exactly the form of $\Gamma_2$ given by Eq.~(\ref{gam}). The condition
$P_3\  \Gamma_1 = 0$ also leads to the $ \Gamma_1$ of Eq.~(\ref{gam}).
For $\Delta_2$ we have:
\begin{equation}
\frac{v_2}{\sqrt{2}} e^{-i \alpha} \Delta_2  = 
W^\dagger_L(12) \ P_3 W_L(12)\ M_u =   P_3 \  M_u
\end{equation}
In this case, the projector $P_3$ picks the diagonal (33) entry of $M_u$,
which together with   $P_3\  \Delta_1 = 0$ leads to the matrix forms of 
Eq.~(\ref{del}). The two other models of the same class, with FCNC in the down
sector are obtained by taking the two other projectors, $P_1$ and $P_2$,
in each case. The three other cases with FCNC in the up sector only, correspond
to:
\begin{eqnarray}
N^0_d =  \frac{v_2}{v_1} \ M_d -  \left( \frac{v_2}{v_1} +  \frac{v_1}{v_2}
\right) \  U_{dL}P_i U^\dagger_{dL} \ M_d \\
N^0_u =   \frac{v_2}{v_1} \ M_u  -  \left( \frac{v_2}{v_1} +  \frac{v_1}{v_2}
\right) \  U_{dL}P_i U^\dagger_{dL} \ M_u 
\end{eqnarray}
with $i=1,2,3$ respectively. The special feature of these six different models 
is the fact that there are WB's 
where the $\Gamma$ and the $\Delta$ matrices have sectors with zero textures 
that do not mix with each other and, as BGL have shown,
these models can be implemented by S-type symmetries. \\

It is also possible to build simple models of MFV type with Higgs
mediated FCNC in both sectors, like the one defined by the 
following equations:
\begin{eqnarray}
N^0_d =  \frac{v_2}{v_1} \ M_d -  \left( \frac{v_2}{v_1} +  \frac{v_1}{v_2}
\right) \  U_{uL}P_i U^\dagger_{uL} \ M_d \\
N^0_u =   \frac{v_2}{v_1} \ M_u  -  \left( \frac{v_2}{v_1} +  \frac{v_1}{v_2}
\right) \  U_{dL}P_i U^\dagger_{dL} \ M_u 
\end{eqnarray}
It is also possible to have MFV models beyond standard NFC \cite{Glashow:1976nt}
but without FCNC, like
\begin{eqnarray}
N^0_d =  \frac{v_2}{v_1} \ M_d -  \left( \frac{v_2}{v_1} +  \frac{v_1}{v_2}
\right) \  U_{dL}P_i U^\dagger_{dL} \ M_d \\
N^0_u =   \frac{v_2}{v_1} \ M_u  -  \left( \frac{v_2}{v_1} +  \frac{v_1}{v_2}
\right) \  U_{uL}P_i U^\dagger_{uL} \ M_u 
\end{eqnarray}
One can also construct more involved MFV models of the BGL type: 
\begin{eqnarray}
N^0_d =  \frac{v_2}{v_1} \ M_d -  \left( \frac{v_2}{v_1} +  \frac{v_1}{v_2}
\right) \  U_{uL}P_i U^\dagger_{uL} \ M_d \\
N^0_u =   \frac{v_2}{v_1} \ M_u  -  \left( \frac{v_2}{v_1} +  \frac{v_1}{v_2}
\right) \  U_{uL}P_j U^\dagger_{uL} \ M_u 
\end{eqnarray}
with $i\neq j$. In all cases the $\Gamma$ and $\Delta$ matrices obey 
relations of the same type as those written in Eqs~(\ref{g1g2}) and
(\ref{d1d2}).  However, the zero texture structure of these models is
more involved than in the BGL case and the question of assuring 
its loop stability, through the introduction of
symmetries, is not obvious \cite{progress}.

\section{Models with Three Higgs Doublets}
Let us now consider the case of three Higgs doublets in the
context of MFV, where the analogous to  Eq.~(\ref{1e2}) 
includes the Yukawa terms of the third Higgs doublet.

After spontaneous symmetry breakdown the Higgs doublets can be
decomposed as:
\begin{eqnarray}
\Phi_j = e^{i\alpha_j}   \left( \begin{array}{c}
\phi^+_j \\
\frac{1}{\sqrt{2}}(v_j + \rho_j + i \eta_j)
\end{array}\right), \qquad j= 1, 2, 3
\end{eqnarray}
with real scalar fields $\rho_j$, $\eta_j$.
The following transformation:
\begin{eqnarray}
\left( \begin{array}{c}
H^0 \\
R \\
R^\prime
\end{array}\right) = O \ \left( \begin{array}{c}
\rho_1 \\
\rho_2\\
\rho_3
\end{array}\right), \qquad \left( \begin{array}{c}
G^0 \\
I \\
I^\prime
\end{array}\right) = O \ \left( \begin{array}{c}
\eta_1 \\
\eta_2\\
\eta_3
\end{array}\right)
\end{eqnarray}
\noindent
with the matrix $O$ given by:
\begin{eqnarray}
O  = \left[\begin{array}{ccc}  
\frac{v_1}{v}  &\frac{v_2}{v}   & \frac{v_3}{v}  \\
\frac{v_2}{v^\prime}  & -\frac{v_1}{v^\prime} & 0  \\
\frac{v_1}{v^{\prime \prime}} & \frac{v_2}{v^{\prime \prime}} & 
\frac{ - (v^2_1 + v^2_2)/v_3}{v^{\prime \prime}} 
\end{array}\right] 
\end{eqnarray}
where $v=\sqrt{v^2_1 + v^2_2 + v^2_3} $, $v^\prime = \sqrt{v^2_1 + v^2_2}$
and $v^{\prime \prime} = \sqrt{v^2_1 + v^2_2 + (v^2_1 + v^2_2)^2 / v^2_3}$.
The orthogonal matrix $O$
singles out $H^0$ and the neutral pseudo-Goldstone boson $G$.
$H^0$ has couplings to the quarks which are proportional
to the mass matrices. In general, flavour changing neutral currents arise
from the couplings to the remaining four neutral Higgs fields.
The diagonalization of the quark mass matrices gives rise to the following
neutral Higgs interactions of the physical quarks:
\begin{eqnarray}
L_Y (neutal) &=&  - 
\frac{H^0}{v} \left(  \bar{d_L} D_d \ d_R + \bar{u_L} D_u u_R  \right) - 
\nonumber \\
& - &  \bar{d_L} \frac{1}{ v\prime } \  {\cal{N}}_d (R+iI) d_R -
 \bar{u_L} \frac{1}{v\prime }  {\cal{N}}_u (R-iI) u_R - \\
& - &  \bar{d_L} \frac{1}{v^{\prime \prime}} \
 {\cal{N}}^\prime_d (R^\prime +iI^\prime ) d_R -
  \bar{u_L} \frac{1}{v^{\prime \prime}} \  
 {\cal{N}}^\prime_u (R^\prime -iI^\prime ) u_R +
\mbox{h.c.}
\nonumber
\end{eqnarray}
with
\begin{eqnarray}
 {\cal{N}}_d & =&  \frac{1}{\sqrt{2}} U^\dagger_{dL}\  
( v_2 e^{i \alpha_1}  \Gamma_1  -
                           v_1 e^{i \alpha_2} \Gamma_2 )\  U_{dR},  \\
 {\cal{N}}_u &= &\frac{1}{\sqrt{2}} U^\dagger_{uL} ( v_2  e^{-i \alpha_1} \Delta_1  -
                           v_1 e^{-i \alpha_2} \Delta_2 )\  U_{uR}, \\
 {\cal{N}}^\prime_d &=&  \frac{1}{\sqrt{2}} U^\dagger_{dL}\  
( v_1  e^{i \alpha_1} \Gamma_1   +
                           v_2 e^{i \alpha_2} \Gamma_2 
+ x  e^{i \alpha_3} \Gamma_3  )\  U_{dR}, \\
 {\cal{N}}^\prime_u &=&  \frac{1}{\sqrt{2}} U^\dagger_{uL}\  
( v_1 e^{-i \alpha_1} \Delta_1  +
                           v_2 e^{-i \alpha_2} \Delta_2 
+ x  e^{-i \alpha_3} \Delta_3  )\  U_{uR}
\end{eqnarray}
where $x = - (v^2_1 + v^2_2)/v_3$. 

For definiteness, let us consider $ {\cal{N}}_d $ and $ {\cal{N}}^\prime_d$,
which can be written:
\begin{eqnarray}
 {\cal{N}}_d &=& \frac{v_2}{v_1} D_d - \frac{v_2}{\sqrt{2}}
\left( \frac{v_2}{v_1} +  \frac{v_1}{v_2}\right)  U^\dagger_{dL}
e^{i \alpha_2} \Gamma_2 \ U_{dR} -  
\frac{v_2 \ v_3}{v_1\sqrt{2}} U^\dagger_{dL}e^{i \alpha_3} \Gamma_3  U_{dR}\\
  {\cal{N}}^\prime_d & = &  D_d - 
\frac{v_3 - x}{\sqrt{2}} \  U^\dagger_{dL}
e^{i \alpha_3} \Gamma_3 \ U_{dR}
\end{eqnarray}

Imposing the following symmetry on the Lagrangian:
\begin{eqnarray}
Q^0_{L1}& \rightarrow & \omega \  Q^0_{L1}, \qquad
Q^0_{L2} \rightarrow \omega^2 \  Q^0_{L2}, \qquad
Q^0_{L3} \rightarrow \omega^4 \  Q^0_{L3}, \nonumber \\
\Phi_1 & \rightarrow & \omega \  \Phi_1, \qquad
\Phi_2 \rightarrow  \omega^2 \ \Phi_2, \qquad
\Phi_3 \rightarrow  \omega^4 \ \Phi_3, \label{qqq} \\
u^0_{R1} & \rightarrow & \omega^2  \ u^0_{R1}, \qquad 
u^0_{R2} \rightarrow \omega^4  \ u^0_{R2}, \qquad
u^0_{R3} \rightarrow \omega^8  \ u^0_{R3},  \nonumber \\
d^0_{Rj} & \rightarrow & d^0_{Rj} \nonumber
\end{eqnarray}
with $ \omega = \exp{i \pi/4}$,
restricts the Yukawa coupling matrices to have the following structure:
\begin{eqnarray}
 \Gamma_1 & = & \left[\begin{array}{ccc}  
\times  & \times & \times \\
0  & 0 &  0 \\
0 & 0 & 0 
\end{array}\right]; \quad
 \Gamma_2   =  \left[\begin{array}{ccc}  
0 & 0 & 0  \\
\times & \times &  \times \\ 
0 & 0 & 0 
\end{array}\right]; \quad
 \Gamma_3   =  \left[\begin{array}{ccc}  
0 & 0 & 0  \\
0 & 0 & 0 \\
\times & \times &  \times 
\end{array}\right] \\
 \Delta_1  & = & \left[\begin{array}{ccc}  
\times  & 0  & 0 \\
0 & 0 &  0 \\
0 & 0 & 0 
\end{array}\right]; \quad 
 \Delta_2  =  \left[\begin{array}{ccc}  
0  & 0 & 0 \\
0 & \times &  0 \\
0 & 0 & 0
\end{array}\right]; \quad
 \Delta_3   =  \left[\begin{array}{ccc}  
0  & 0 & 0 \\
0 & 0 &  0 \\
0 & 0 & \times
\end{array}\right] 
\end{eqnarray}
where $\times$ denotes an arbitrary entry while the zeros are imposed
by the above symmetry. 

It can be readily verified that in this case there are Higgs mediated FCNC
only in the down sector, with $ {\cal{N}}_d $ and $ {\cal{N}}^\prime_d$
given by:
\begin{eqnarray}
 ({\cal{N}}_d)_{ij} &=& \frac{v_2}{v_1} (D_d)_{ij} - 
\left( \frac{v_2}{v_1} +  \frac{v_1}{v_2}\right)
(V^\dagger_{CKM})_{i2} (V_{CKM})_{2j} (D_d)_{jj} - \nonumber \\
&-&  \frac{v_2}{v_1}
(V^\dagger_{CKM})_{i3} (V_{CKM})_{3j} (D_d)_{jj} \\
({\cal{N}}^\prime_d)_{ij} & = &  (D_d)_{jj} - 
\frac{v_3 -x}{v_3} (V^\dagger_{CKM})_{i3} (V_{CKM})_{3j} (D_d)_{jj} 
\end{eqnarray}
In this case the couplings  ${\cal{N}}_d $ include terms that
violate flavour proportional to \\
$(V^\dagger_{CKM})_{i2} (V_{CKM})_{2j} (D_d)_{jj}$ together with 
terms proportional to \\
$(V^\dagger_{CKM})_{i3} (V_{CKM})_{3j} (D_d)_{jj}$. The couplings
$ {\cal{N}}^\prime_d$ only include terms that violate flavour proportional
to 
$(V^\dagger_{CKM})_{i3} (V_{CKM})_{3j} (D_d)_{jj}$. It is clear
that all Higgs mediated neutral couplings are only function of $V_{CKM}$
and terefore the symmetry of Eq.~(\ref{qqq}) leads to a MFV structure 
in the context of a three Higgs-doublet model. From a phenomenological 
point of view, there is an important difference between the scalar
FCNC in this MFV three Higgs doublet model and those encountered 
in the MFV two Higgs doublet model considered in the previous
chapters. In the case of two Higgs doublet models, there is one 
variant of the BGL models where the tree level Higgs mediated $\Delta S=2$ 
amplitude is naturally suppressed by terms proportional to 
$(V^*_{td} V_{ts})^2$. This very strong suppression opens the possibility
of having neutral Higgs relatively light of order $10^2$ Gev,
without entering in conflict with the size of the $K_L-K_S$ mass difference
or the strength of CP violation in the kaon sector. In the case
of the MFV three Higgs doublet model $ {\cal{N}}_d $ includes FCNC
terms where the suppression factor in $\Delta S=2$ transitions is only 
$(V^*_{cd} V_{cs})^2$,  which then requires quite heavy neutral Higgs,
with mass of order Tev.

\section{Conclusions}
We have analysed how to extend the MFV concept to general 
multi-Higgs models without NFC in the Higgs sector. We have 
studied in special detail the case of two Higgs doublet models, 
analysing the requirements which have to be satisfied in order
that the neutral Higgs couplings to quarks be only functions
of $V_{CKM}$, with no other flavour dependent parameters. The
Branco-Grimus-Lavoura (BGL) models proposed some time ago are an 
example where the MFV constraints are satisfied as the result
of a symmetry of the Lagrangian. We have proposed a general 
MFV expansion of the neutral Higgs couplings to quarks and 
have shown that the BGL models correspond to specific values 
of the coefficients of the proposed MFV expansion and,
in addition, we have shown that the values of these coefficients are 
fixed by the symmetry.

Multi-Higgs models with Higgs mediated FCNC have a rich 
phenomenology and some of its aspects have been recently 
analysed in the literature \cite{quatro}. A detailed phenomenological
analysis of multi-Higgs MFV models without NFC, is beyond the scope of 
this paper and will be left to a separate work \cite{more}

\section*{Acknowledgements}

This work was partially supported by Funda\c c\~ ao para a Ci\^ encia
e a Tecnologia (FCT, Portugal) through the projects
CERN/FP/83503/2008 and CFTP-FCT Unit 777 which are partially funded
through POCTI (FEDER),  by Marie Curie RTN
MRTN-CT-2006-035505, by  Accion Complementaria Luso-Espanhola
PORT2008--03,
by European FEDER, Spanish  MICINN under grant FPA2008--02878  .
GCB and MNR are very grateful for the hospitality of Universisitat de
Val\` encia during their visits. FJB is very grateful for
the hospitality of CFTP/IST Lisbon during his visits.

\end{document}